\documentclass[12pt,reqno]{article}

\usepackage{amssymb, amsmath}
\usepackage{amsthm}
\usepackage{amsfonts}
\usepackage{latexsym}
\usepackage{graphicx}
\usepackage{epsf}

\newcommand{\N}{\mathbb{N}}
\newcommand{\Nplus}{\mathbb{N}^{+}}

\begin{document}

\theoremstyle{plain}
\newtheorem{theorem}{Theorem}
\newtheorem{lemma}[theorem]{Lemma}
\newtheorem{corollary}[theorem]{Corollary}

\theoremstyle{definition}
\newtheorem{definition}[theorem]{Definition}
\newtheorem{example}[theorem]{Example}

\begin{center}
{\LARGE\bf Fast Computing the Algebraic Degree of Boolean Functions\\
}
\vskip 5mm
{\large Valentin Bakoev}\\
\vskip 2mm
Faculty of Mathematics and Informatics,\\
``St. Cyril and St. Methodius'' University of Veliko Tarnovo,\\
2 Theodosi Tarnovski Str., 5003 Veliko Tarnovo, Bulgaria;\\
email: v.bakoev@ts.uni-vt.bg
\end{center}

\vskip .2 in
\begin{abstract}
	Here we consider an approach for fast computing the algebraic degree of Boolean functions. It combines fast computing the ANF (known as ANF transform) and thereafter the algebraic degree by using the weight-lexicographic order (WLO) of the vectors of the $n$-dimensional Boolean cube. Byte-wise and bitwise versions of a search based on the WLO and their implementations are discussed. They are compared with the usual exhaustive search applied in computing the algebraic degree. For Boolean functions of $n$ variables, the bitwise implementation of the search by WLO has total time complexity $O(n.2^n)$. When such a function is given by its truth table vector and its algebraic degree is computed by the bitwise versions of the algorithms discussed, the total time complexity is $\Theta((9n-2).2^{n-7})=\Theta(n.2^n)$. All algorithms discussed have time complexities of the same type, but with big differences in the constants hidden in the $\Theta$-notation. The experimental results after numerous tests confirm the theoretical results---the running times of the bitwise implementation are dozens of times better than the running times of the byte-wise algorithms. 
\end{abstract}
\vspace{1 mm}

\vspace{1 mm}
\section{Introduction}
\label{Intro}
	Boolean functions are of great importance in the modern cryptography, coding theory, digital circuit theory, etc. When they are used in the design of block ciphers, pseudo-random numbers generators (PRNG) in stream ciphers etc., they should satisfy certain cryptographic criteria \cite{CC_BFCECC, CC_VBFC, ACANT}. One of the most important cryptographic parameters is the \textbf{\textit{algebraic degree}} of a Boolean function or vectorial Boolean function, called also an S-box. This degree should be higher in order the corresponding Boolean function (or S-box, or PRNG) to be resistant to various types of cryptanalytic attacks. The process of generating such Boolean functions needs this parameter, as well as the other important cryptographic parameters, to be computed as fast as possible. In this way, more Boolean functions can be generated and a better choice among them can be done.

	Let $f$ be a Boolean function of $n$ variables given by its Truth Table vector denoted by $TT(f)$. There are \textbf{two main approaches} for computing the algebraic degree of $f$. The \textbf{first} one uses the Algebraic Normal Form (ANF) representation of $f$ and selects the monomial of the highest degree in it. The \textbf{second} approach uses only the $TT(f)$, its weight, support, etc., without computing the ANF of $f$. In \cite{MWNS, CC_BFCECC, ACANT, CGV} it is proven that if $TT(f)$ has an odd weight, then the algebraic degree of $f$ is maximal. This condition holds for the half of all Boolean functions and it can be verified very easily. The algorithms proposed in \cite{CGV} work only with the $TT(f)$ and use this property. They are fast for just over half of all Boolean functions of $n$ variables. However, when these algorithms are compared with an algorithm of the first type (i.e., based on ANF), the computational results set some questions about the efficiency of algorithms used for computing the ANF and thereafter the algebraic degree. This is one of the reasons that motivated us to do a more comprehensive study of the first approach---fast computing the algebraic degree of Boolean functions by their ANFs. We have already done three basic steps in this direction discussed in sections \ref{Prelim_Res} and \ref{BitwiseAppr}. Here we represent the next step which is a natural continuation of the previous ones. It includes a bitwise implementation of the ANF Transform (ANFT) followed by a bitwise computing the algebraic degree by using masks for one special sequence representing the weight-lexicographic order (WLO) of the vectors of Boolean cube. 
	
	The paper is structured as follows. The basic notions are given in Section \ref{Bas_Not}. In Section \ref{Prelim_Res} we outline some preliminary results about the enumeration and distribution of Boolean functions of $n$ variables according to their algebraic degrees, as well as the WLO of the vectors of the Boolean cube and the corresponding sequences. At the beginning of Section \ref{AD_by_WLO}, an algorithm for computing the algebraic degree of Boolean function by using the WLO sequence is discussed. Section \ref{BitwiseAppr} starts with a comment on the preliminary results about the bitwise ANF transform. Thereafter, a search by using masks for the WLO sequence is considered. Section \ref{Exp_Res} shows a scheme of computations and used algorithms. The time complexities of the algorithms under consideration are summarized and the experimental results after numerous tests are given. They are used for comparison of the byte-wise and bitwise implementations of the proposed algorithms. The \textbf{\textit{general conclusion}} is: in computing the algebraic degree of a Boolean function it is worth to use the bitwise implementation of proposed algorithms instead of the byte-wise one---it is tens of times faster. In the last section, some ideas about the forthcoming steps of this study are outlined. Experiments in one of these directions have already begun and their first results are good.
\section{Basic Notions}
\label{Bas_Not}
Here $\N$ denotes the set of natural numbers. We consider that $0\in \N$ and
$\N^{+}=\N \backslash \{0\}$ is the set of positive natural numbers.

Usually, the \textit{$n$-dimensional Boolean cube} is defined as $\{0,1\}^n=$ $ \{(x_1,x_2,\dots,x_n)|$ $x_i\in \{0,1\}, \forall\, i=1,2,\dots, n \}$, i.e., it is the set of all $n$-dimensional binary vectors. So $|\{0,1\}^n| = |\{0,1\}|^n= 2^n$. Further, we use the following alternative, inductive and constructive definition.
\begin{definition}
\label{D10}
\normalfont{
	1) The set $\{0,1\}= \{(0),(1)\}$ is called \textit{one-dimensional 
Boolean cube} and its elements $(0)$ and $(1)$ are called \textit{one-dimensional binary vectors}.

	2) Let $\{0,1\}^{n-1}=\{ \alpha_0, \alpha_1, \dots$, $\alpha_{2^{n-1}-1} \}
$ be the \textit{$(n-1)$-dimensional Boolean cube} and $\alpha_0, \alpha_1, 
\dots, \alpha_{2^{n-1}-1}$ be its \textit{$(n-1)$-dimensional binary vectors}. 

	3) The \textit{$n$-dimensional Boolean cube} $\{0,1\}^n$ is built by taking 
the vectors of $\{0,1\}^{n-1}$ twice: firstly, each vector of $\{0,1\}^{n-1}$ is prefixed by zero, and thereafter each vector of $\{0,1\}^{n-1}$ is prefixed by one:
\begin{eqnarray*}
\{0,1\}^n= \{(0,\alpha_0),(0,\alpha_1), \dots ,(0,\alpha_{2^{n-1}-1}), \\                 (1,\alpha_0),(1,\alpha_1), \dots ,(1,\alpha_{2^{n-1}-1}) \}.
\end{eqnarray*}
}
\end{definition}

	For an arbitrary vector $\alpha=(a_1, a_2, \dots, a_n)\in \{0,1\}^n$, 
the natural number $\#\alpha=\sum_{i=1}^n a_i.2^{n-i}$ is called a  \textit{serial number} of the vector $\alpha$. So $\#\alpha$ is the natural number having $n$-digit binary representation $a_1 a_2 \dots a_n$. A \textit{(Hamming) weight} of $\alpha$ is the natural number $wt(\alpha)$, equal to the number of non-zero coordinates of $\alpha$, i.e., $wt(\alpha)= \sum_{i=1}^n a_i$. For any $k\in \N$, $k\leq n$, the set of all $n$-dimensional binary vectors of weight $k$ is called a \textit{$k$-th layer} of the $n$-dimensional Boolean cube. It is denoted by $L_{n,k}= \{\alpha|\, \alpha \in \{0,1\}^n : wt(\alpha)=k\}$ and we have $|L_{n,k}|= \binom{n}{k}$, for $k=0, 1, \dots, n$. These numbers are the binomial coefficients from the $n$-th row of Pascal's triangle and so $\sum_{k=0}^n \binom{n}{k}= 2^n= |\{0, 1\}^n|$. The family of all layers $L_n=\{ L_{n,0}, L_{n,1},\dots,L_{n,n}\}$ is a \textit{partition} of the $n$-dimensional Boolean cube into layers. 

	For arbitrary vectors $\alpha= (a_1, a_2, \dots, a_n)$ and $\beta= (b_1, b_2,$ $\dots, b_n) \in \{0,1\}^n$, we say that "$\alpha$ \textit{precedes lexicographically} $\beta$" and denote this by $\alpha\leq \beta$, if $\alpha=\beta$ or if $\exists\, k, 1\leq k\leq n$, such that $a_k<b_k$ and $a_i=b_i$, for all $i<k$. The relation "$\leq$" is a \textit{total} (unique) order in $\{0,1\}^n$, called \textit{lexicographic order}. The vectors of $\{0,1\}^n$ are ordered lexicographically in the sequence $\alpha_0, \alpha_1, \dots \alpha_k, \dots, \alpha_{2^n-1}$ if and only if: 
\begin{itemize}
	\item $\alpha_l\leq \alpha_k, \forall\, l\leq k$ and $\alpha_k\leq \alpha_r, \forall\, k\leq r$;
	\item the sequence of their serial numbers $\#\alpha_0, \#\alpha_1,\dots,$ $\#\alpha_k,\dots, \#\alpha_{2^n-1}$ is exactly $0, 1,$ $\dots, k,\dots, 2^n-1$.
\end{itemize}
	
	A \textit{Boolean function} of $n$ variables (denoted usually by $x_1, x_2, \dots , x_n$) is a mapping $f:\{0,1\}^n \rightarrow \{0,1\}$, i.e. $f$ maps any binary input $x=(x_1, x_2,\dots , x_n) \in \{0,1\}^n$ to a single binary output $y = f(x)\in \{0,1\}$. Any Boolean function $f$ can be represented in a unique way by the vector of its functional values, called a \textit{Truth Table} vector and denoted by $TT(f)= (f_0, f_1, \dots f_{2^n-1})$, where $f_i= f(\alpha_i)$ and $\alpha_i$ is the $i$-th vector in the lexicographic order of $\{0,1\}^n$, for $i= 0, 1, \dots, 2^n-1$.	The set of all Boolean functions of $n$ variables is denoted by ${\cal{B}}_n$ and its size is $|{\cal{B}}_n|= 2^{2^n}$.

	Another unique representation of the Boolean function $f\in {\cal{B}}_n$ is the \textit{algebraic normal form} (ANF) of $f$, which is a multivariate polynomial
\begin{eqnarray*}
\label{F10}
f(x_1, x_2, \dots, x_n)= \bigoplus_{\gamma\in \{0,1\}^n} a_{\#\gamma}\, x^{\gamma}\,.
\end{eqnarray*}
	
	Here $\gamma=(c_1, c_2, \dots, c_n)\in \{0,1\}^n$, the coefficient $a_{\#\gamma}\in \{0, 1\}$, and $x^{\gamma}$ means the monomial $x_1^{c_1} x_2^{c_2} \dots x_n^{c_n}= \prod_{i=1}^n x_i^{c_i}$, where $x_i^0=1$ and $x_i^1=x_i$, for $i=1, 2, \dots n$. A \textit{degree} of the monomial $x=x_1^{c_1} x_2^{c_2} \dots x_n^{c_n}$ is the integer $deg(x)= wt(\gamma)$---it is the number of variables of the type $x_i^1=x_i$, or the essential variables for $x^\gamma$. The \textit{algebraic degree} (or simply \textit{degree}) of $f$ is defined as $deg(f)= max\{deg(x^{\gamma})|\, a_{\#\gamma}=1\}$. When $f\in {\cal{B}}_n$ and the $TT(f)$ is given, the values of the coefficients $a_0, a_1, \dots, a_{2^n-1}$ can be computed by a fast algorithm, usually called an \textit{ANF transform} (ANFT)\footnote{In dependence of the area of consideration, the same algorithm is called also (fast) M\"obius Transform, Zhegalkin Transform, Positive polarity Reed-Muller Transform, etc.}. The ANFT is well studied, it is derived in different ways by many authors, for example \cite{CC_BFCECC, ACANT, AJOUX}. Its byte-wise implementation has a time-complexity $\Theta(n.2^n)$. The vector $(a_0, a_1, \dots, a_{2^n-1})\in \{0,1\}^n$ obtained after the ANFT is denoted by $A_f$. When $f\in {\cal{B}}_n$ is the constant zero function (i.e., $TT(f)=(0,0,\dots, 0)$), its ANF is $A_f= (0,0,\dots,0)$ and its algebraic degree is defined as $deg(f)= -\infty$. If $f$ is the constant one function ($TT(f)=(1,1,\dots, 1)$), then $A_f= (1,0,0,\dots,0)$ and $deg(f)=0$. 
\section{Some Preliminary Results}
\label{Prelim_Res}
\subsection{Distribution of Boolean Functions According to their Algebraic Degrees}
\label{Distrib}
	It is well-known that half of all Boolean functions of $n$ variables have an algebraic degree equal to $n$, for $n\in\Nplus$ \cite{MWNS, CC_BFCECC, ACANT, CGV}. Furthermore, in \cite[p. 49]{CC_BFCECC} Carlet notes that when $n$ tends to infinity, random Boolean functions have almost surely algebraic degrees at least $n-1$. We consider that the overall enumeration and distribution of all Boolean functions of $n$ variables ($n\in\Nplus$) according to their algebraic degrees is very important for our study. The paper where we explore them is still in review, but some results can be seen in OEIS \cite{OEIS}, sequence A319511. We will briefly outline the results needed for further exposition.
	
	Let $d(n,k)$ be the number of all Boolean functions $f \in {\cal{B}}_n$ such that $deg(f)=k$.
\begin{theorem}
\label{Th10} For arbitrary integers $n\in \N$ and $0\leq k\leq n$, the number
	\begin{eqnarray*}
\label{F20}	
	d(n, k)= (2^{\binom{n}{k}}-1).2^{\sum_{i=0}^{k-1} \binom{n}{i}}\,.
\end{eqnarray*}
\end{theorem}
Sketch of proof: let $X$ be the set of $n$ variables. There are $\binom{n}{k}$ monomials of degree $=k$ because so many are the ways to choose $k$ variables from $X$. The first multiplier in the formula denotes the number of ways to choose at least one such monomial to participate in the ANF. The second multiplier is the number of ways to choose 0 or more monomials of degrees $<k$ and to add them to the ANF.

\begin{corollary}
\label{Cor10}
The number $d(n,n-1)$ tends to $\displaystyle \frac{1}{2}\cdot|{\cal{B}}_n|$ when $n\rightarrow\infty$.
\end{corollary}
	
	Let $p(n,k)$ be the discrete probability a random Boolean function $f \in {\cal{B}}_n$ to have an algebraic degree $=k$. It is defined as
\[
p(n,k)=\frac{d(n,k)}{|{\cal{B}}_n|}=\frac{d(n,k)}{2^{2^n}},
\]
for $n\geq 0$ and $0\leq k\leq n$. The values of $p(n,k)$ obtained for a fixed $n$ give the distribution of the functions from ${\cal{B}}_n$ according to their algebraic degrees. Table~\ref{Ta20:distribution} represents this distribution, for $3\leq n\leq 10$ and $n-3\leq k \leq n$. The values of $p(n,k)$ in it are rounded up to 10 digits after the decimal point.
Furthermore, $p(n,k)\approx 0$, for $0\leq k<n-3$, and their values are not shown in the table.
\begin{table}[ht]
\caption{Distribution of the functions from ${\cal{B}}_n$ according to their algebraic degrees, for $n= 3,4,\dots, 10$}%%
\label{Ta20:distribution}
\centering
{\small
	\begin{tabular}{|r|l|l|l|l|}
	\hline
  & \multicolumn{4}{c |}{The values of $p(n,k)$, for:} \\
	  \cline{2-5}
	$n\,$ & \multicolumn{1}{c |}{$k=n-3$} &  \multicolumn{1}{c |}{$k=n-2$} &  			\multicolumn{1}{c |}{$k=n-1$} & \multicolumn{1}{c |}{$k=n$}\\
	\hline
	3 & 0.00390625    & 0.0546875    & 0.4375 & 0.5 \\
	\hline
	4  & 0.0004577637 & 0.0307617187 & 0.46875 & 0.5 \\
	\hline
	5  & 0.0000152439 & 0.0156097412 & 0.484375 & 0.5 \\
	\hline
	6  & 0.0000002384 & 0.0078122616 & 0.4921875 & 0.5 \\
	\hline
	7  & 0.0000000019 & 0.0039062481 & 0.49609375 & 0.5 \\
	\hline
	8  & 0            & 0.0019531250 & 0.498046875 & 0.5 \\
	\hline
	9  & 0            & 0.0009765625 & 0.4990234375 & 0.5 \\
	\hline
	10 & 0            & 0.0004882812 & 0.4995117187 & 0.5 \\
	\hline
	\end{tabular}
}
\end{table}	

These results were used:
\begin{itemize}
	\item To check for representativeness the files used to test all  algorithms discussed here. These are 4 files containing $10^6, 10^7, 10^8$ and $10^9$ randomly generated unsigned integers in 64-bit computer words. We used each of these files as an input for Boolean functions of $6, 8, 10, \dots, 16$ variables (reading $2^{n-6}$ integers from the chosen file) and we computed the algebraic degrees of all these functions. The absolute value of the difference between the theoretical and computed distribution is less than $0.88\%$ (it exceeds $0.1\%$ in only a few cases), for all tests. So we consider that the algorithms work with samples of Boolean functions which are representative enough.
	\item When creating the algorithms represented in the following sections. The distribution shows why the WLO has been studied in detail and what to expect for the running time of algorithms that use WLO.
\end{itemize}
\subsection{WLO of the Vectors of $n$-dimensional Boolean Cube}
	The simplest algorithm for computing the algebraic degree of a Boolean function is an \textit{Exhaustive Search} (we refer to it as \textbf{\textit{ES algorithm}}): if $f\in{\cal{B}}_n$ and $A_f=(a_0, a_1$, $\dots, a_{2^n-1})$ is given, it checks consecutively whether $a_i=1$, for $i=0,1,\dots, 2^n-1$. The algorithm selects the vector of maximal weight among all vectors $\alpha_i\in \{0,1\}^n$ such that $a_i=1$. The algorithm checks exhaustively all values in $A_f$ (which correspond to the lexicographic order of the vectors of $\{0,1\}^n$) and so it performs $\Theta(2^n)$ checks.
	
	The basic parts of a faster way for the same computing are considered in \cite{VB-WLO-S, VB-WLO-A}. Here they are given in short, but all related notions, proofs, illustrations, algorithms and programming codes, details, etc., can be seen in \cite{VB-WLO-A}.
	
	The \textit{sequence of layers} $L_{n,0}, L_{n,1}, \dots, L_{n,n}$ gives an \textit{order} of the vectors of $\{0,1\}^n$ in accordance with their weights. When $\alpha, \beta \in \{0,1\}^n$ and $wt(\alpha) < wt(\beta)$, then $\alpha$ \textit{precedes} $\beta$ in the sequence of layers, and if $wt(\alpha)=wt(\beta)=k$, then $\alpha, \beta \in L_{n,k}$ and there is no precedence between them. We define the corresponding relation $R_{<_{wt}}$ as follows: for arbitrary $\alpha, \beta \in \{0, 1\}^n$, $(\alpha, \beta) \in R_{<_{wt}}$ if $wt(\alpha) < wt(\beta)$ or if $\alpha=\beta$. When $(\alpha, \beta) \in R_{<_{wt}}$ we say that "$\alpha$ \textit{precedes by weight} $\beta$" and write also $\alpha <_{wt} \beta$. Thus $R_{<_{wt}}$ is a partial order in $\{0, 1\}^n$ and we refer to it (and to the order determined by it) as a \textit{Weight-Order} (WO). To develop an algorithm we use the serial numbers of the vectors in the sequence of layers instead of the vectors themselves. For an arbitrary layer $L_{n,k}=\{\alpha_0, \alpha_1, \dots, \alpha_m\}$ of $\{0,1\}^n$, we define the \textit{sequence of serial numbers} of the vectors of $L_{n,k}$ and denote it by $l_{n,k}=\#\alpha_0, \#\alpha_1, \dots, \#\alpha_m$. Let $l_n= l_{n,0}, l_{n,1}, \dots, l_{n,n}$ be the \textit{sequence of all serial numbers} corresponding to the vectors in the sequence of layers $L_{n,0}, L_{n,1}, \dots, L_{n,n}$. Thus $l_n$ represents a WO of the vectors of $\{0,1\}^n$ and we call $l_n$ a \textit{WO sequence} of $\{0,1\}^n$. One of all possible $\prod_{k=0}^n \binom{n}{k}!$ WO sequences\footnote{You can see the sequence  A051459 in the OEIS \cite{OEIS} for details.} deserves a special attention. Firstly, we define the operation \textit{addition of the natural number to a sequence} as follows: if $n, m\in \N^+$ and $s= a_1, a_2, \dots, a_n$ is a sequence of integers, then  $s+m= a_1+m, a_2+m, \dots, a_n+m$. Following Definition \ref{D10}, we obtain:
\begin{definition}
\label{D20}
\normalfont{
	1) The WO sequence of the one-dimensional Boolean cube is $l_1= 0, 1$.
	
	2) Let $l_{n-1}= l_{n-1,0}, l_{n-1,1}, \dots, l_{n-1,n-1}$ be the WO sequence of the $(n-1)$-dimen-sional Boolean cube.	

	3) The WO sequence of $n$-dimensional Boolean cube $l_n= l_{n,0}, l_{n,1}, \dots, l_{n,n}$ is defined as follows: 
	
	$\bullet$ $l_{n,0}= 0$ and it corresponds to the layer $L_{n,0}= \{\tilde{0}_n\}$, where $\tilde{0}_n$ is the zero vector of $n$ coordinates; 
	
	$\bullet$	$l_{n,n}= 2^n-1$ and it corresponds to the layer $L_{n,n}= \{\tilde{1}_n\}$, where $\tilde{1}_n$ is the all-ones vector of $n$ coordinates;
	
	$\bullet$	$l_{n,k}= l_{n-1,k},\, l_{n-1,k-1}+2^{n-1}$, for $k=1, 2, \dots, n-1$. Here $l_{n,k}$ is a concatenation of two sequences: the sequence $l_{n-1,k}$ is taken (or copied) firstly, and the sequence $l_{n-1,k-1} + 2^{n-1}$ follows after it. The sequence $l_{n,k}$ corresponds to the layer $L_{n,k}$.
}
\end{definition}
\begin{theorem}
\label{T10}
	Let $n\in \Nplus$ and $l_n= l_{n,0}, l_{n,1}, \dots, l_{n,n}$ be the WO sequence, obtained in accordance with Definition \ref{D20}. Then, the serial numbers in the sequence $l_{n,k}$ determine a lexicographic order of the vectors of the corresponding layer $L_{n,k}$, for $k=0,1,\dots, n$.
\end{theorem}

	Theorem \ref{T10} is proven by mathematical induction in \cite{VB-WLO-A}. It states that Definition \ref{D20} determines a \textit{second criterion} for ordering the vectors within the existing WO of the Boolean cube---this is the \textit{lexicographic order}. Since it is a total order for each subsequence $l_{n,k}$, $0\leq k\leq n$, a total weight order for the sequence $l_n$ is obtained. We call it a \textit{Weight-Lexicographic Order} (WLO).

	The \textit{WLO algorithm} is based on Definition \ref{D20} and Theorem \ref{T10}, and so they imply its correctness. For a given input $n\in \Nplus$, it starts from $l_1$ and computes consecutively the sequences $l_2, l_3, \dots, l_n$. Some results computed by the algorithm are given in  Table \ref{Tab:Results-WLO-Alg}. More results can be seen in OEIS \cite{OEIS}, sequence A294648. 
\begin{table}[!ht]
	\caption{Results obtained by the WLO algorithm for $n=1,2,\dots, 5$}
	\label{Tab:Results-WLO-Alg}
	\begin{center}
		\begin{tabular}{|c|l|}
		\hline
		$\, n\, $ &\, $l_n$\\
		\hline
		1  &\, 0, 1 \\
		\hline
    2  &\, 0, 1, 2, 3 \\
    \hline
    3  &\, 0, 1, 2, 4, 3, 5, 6, 7 \\
    \hline
    4  &\, 0, 1, 2, 4, 8, 3, 5, 6, 9, 10, 12, 7, 11, 13, 14, 15 \\
    \hline
    5  &\, 0, 1, 2, 4, 8, 16, 3, 5, 6, 9, 10, 12, 17, 18, 20, 24, 7, 11, 13, 14, 19, 21, 22, 25, $\dots$\\
		\hline
		\end{tabular}
	\end{center}
\end{table}

	The time complexity of the WLO algorithm is $\Theta(2^n)$, it is  exponential with respect to the size of the input $n$. Furthermore, it is linear  with respect to the size of the output. The space complexity of the algorithm is of the same type. We note that the running time for precomputation of the sequence $l_n$ in a lookup table is negligible ($\approx 0$ seconds).
\section{Computing the Algebraic Degree of Boolean Functions by WLO}
\label{AD_by_WLO}
\subsection{Byte-wise Approach}
	The terms of the WLO sequence $l_n$ form a permutation of the numbers $0, 1, \dots, 2^n-1$ and we denote this permutation as $l_n=(i_0, i_1, \dots, i_{2^n-1})$. We use the sequence $l_n$ to compute the algebraic degree of a given Boolean function $f\in{\cal{B}}_n$. The proposed algorithm is similar to the ES algorithm, but it checks the coordinates of $A_f=(a_0, a_1, \dots, a_{2^n-1})$ in accordance with the values of $l_n$, from right to left. It starts with the $i_{2^n-1}$-th coordinate of $A_f$. If it is equal to zero the algorithm checks the $i_{2^n-2}$-th coordinate of $A_f$, and so on, looking for the first coordinate of $A_f$ which is equal to one and then it stops. If there is not such a coordinate, then $f$ is the constant zero function. Otherwise, if the algorithm stops the searching on the $i_j$-th coordinate ($0\leq j<2^n$) of $A_f$, it returns the number of the subsequence that contains the number $i_j$ as an output. If $i_j$ is a term of $l_{n,k}$, $0\leq k \leq n$, then the layer $L_{n,k}$ contains a vector which serial number is $i_j$ and therefore $deg(f)=k$. The algorithm is correct, since it follows the WLO and stops at the right place---if it continues with the checks, it will find possible monomials of degree $\leq k$. Thus the algorithm performs $O(2^n)$ checks and this is its \textit{time complexity}. This general estimation concerns a very small number of functions $f\in{\cal{B}}_n$ because the computing will finish after $O(n)$ checks at almost $100\%$ of all such functions (especially when $n$ grows)---as it is shown in Section \ref{Distrib}. Since this algorithm works in a byte-wise manner and after the byte-wise ANFT, we call it \textbf{\textit{Byte-wise WLO algorithm}}. 
\subsection{Bitwise Approach}
\label{BitwiseAppr}
	In \cite{VB-FastANFT} we represented a comprehensive study of the bitwise implementation of the ANFT. When 64-bit computer words are used, the obtained algorithm has a time-complexity $\Theta((9n-2).2^{n-7})$ and a space complexity $\Theta(2^{n-6})$, i.e., both are of the type $\Theta(2^n)$. But the experimental results show that the bitwise version of the algorithm is about 25 times faster in comparison to the byte-wise version\footnote{Both algorithms have been implemented as C++ programs in Code::Blocks 13.12 IDE, built  as 32-bit applications in Release mode and tested with the largest file of $10^9$ integers.}. Analogous research concerning the parallel bitwise implementation of the ANFT is represented in \cite{DBIB} and similar results about its efficiency are obtained.

 	After these results it is natural to think about a bitwise implementation of the last algorithm. Otherwise, bitwise computing an ANFT seems unnecessary, since computing the other cryptographic parameters of Boolean functions needs a byte-wise representation (see Fig. \ref{fig:Scheme}). Our first idea is to check all vectors in the same layer in one (or several) step(s). For this purpose we use $n+1$ masks $m_{n,0}, m_{n,1},\dots, m_{n,n}$ corresponding to the vectors in the layers $L_{n,0}, L_{n,1},\dots, L_{n,n}$. The mask $m_{n,i}$ is a binary vector of the same length as $A_f$ and $m_{n,i}$ contains units only in these bits, whose coordinates correspond to the numbers in the subsequence $l_{n,i}$, for $i=0,1,\dots,n$. So we need to repeat bitwise conjunctions between $A_f$ and $m_{n,i}$, for $i=n, n-1,\dots, 0$, until $A_f\wedge m_{n,i}=0$. If this equality holds for all values of $i$, then $f$ is the constant zero function. Otherwise, if $k$ is the first value of $i$ (when $i$ decreases from $n$ to $0$) such that $A_f\wedge m_{n,k}>0$, then $k$ is the algebraic degree of $f$. So the algorithm stops and returns k. We call it \textbf{\textit{Bitwise WLO algorithm}} accepting that it always uses masks. 

	When $A_f$ occupies one computer word, the algorithm performs at most $n+1$ steps and so its time complexity is $O(n)$, i.e., it is of logarithmic type ($n=\log_2{2^n}$) with respect to the size of the input. If the size of the computer word is $64=2^6$ bits and $f$ is a function of $n>6$ variables, then $TT(f)$ and $A_f$ occupy $s=2^n/64=2^{n-6}$ computer words. So $m_{n,i}$ will occupy $s$ computer words too and the computing $A_f\wedge m_{n,i}$ will be done in $s$ steps, for $i=n, n-1, \dots, 0$. If on some of these steps the conjunction between the corresponding computer words of $A_f$ and $m_{n,i}$ is greater than zero, the algorithm returns $i$ and stops. Therefore, in the general case, the bitwise WLO algorithm has a \textit{time complexity} $O(n+1).O(s)= O(n.2^{n-6})$. This estimation concerns a very small number of functions $f\in{\cal{B}}_n$ again---the computing will finish after $O(1+s)= O(2^{n-6})$ checks at almost $100\%$ of all such functions. 

	Let us consider the masks' generating. For arbitrary $i, 0\leq i \leq n$, it is easy to put units in all these bits of $m_{n,i}$ that correspond to the numbers in the subsequence $l_{n,i}$. We note that we use the serial numbers of the masks, stored in the necessary number of 64-bit computer words, as well as the vectors $TT(f)$ and $A_f$. Furthermore, we  generate them in accordance with the following definition.
\begin{definition}
\label{D30}
\normalfont{
1) For $n=1$, the serial numbers of the masks corresponding to the subsequences $l_{1,0}$ and $l_{1,1}$ are $\#m_{1,0}=2$ and $\#m_{1,1}=1$.

2) Let $\#m_{n-1,0}, \#m_{n-1,1}, \dots, \#m_{n-1,n-1}$ be the serial numbers of the masks corresponding to the subsequences $l_{n-1,0}, l_{n-1,1}, \dots, l_{n-1,n-1}$. 

3) The serial number of the mask $m_{n,i}$ corresponding to the subsequence $l_{n,i}$ is:
\begin{eqnarray*}
\#m_{n,i}=\left\{
\begin{array} {ll}
2^{2^{n-1}}.\#m_{n-1,0}=2^{2^n-1}, \textrm{\ if\ } i=0\,,\\
1, \textrm{\ if\ } i=n\,,\\
2^{2^{n-1}}.\#m_{n-1,i} + \#m_{n-1,i-1}, \textrm{\ if\ } 0<i<n\,,
\end{array}
\right.
\end{eqnarray*} 
for $i=0,1, \dots ,n$.
}
\end{definition}
	
	Definition \ref{D30} corresponds to definitions \ref{D10} and \ref{D20}. Its correctness can be proven strictly by mathematical induction on $n$. The running time for generating (precomputation of) the masks in accordance with Definition \ref{D30} is negligible ($\approx 0$ seconds). We note that when $n>6$, the generating algorithm has some particularities because it works with $s=2^{n-6}$ computer words for each mask. The serial numbers of masks grow exponentially---see Table \ref{Ta2:Masks5vars}, as well as the sequence A305860 in OEIS \cite{OEIS}.
\begin{table}[ht]
  \caption{Serial numbers of the masks, for $n=1,\dots ,5$}
  \label{Ta2:Masks5vars}
  \centering
  \small{
  \begin{tabular}{|c|r|r|r|r|r|r|}
  \hline
	\,$n$\, & $\#m_{n,0}$ &  $\#m_{n,1}$ &  $\#m_{n,2}$ &  $\#m_{n,3}$ &  $\#m_{n,4}$ & $\#m_{n,5}$\\
	\hline
	1 &          2 &          1 &    --     &    --    & --    & --\\
	\hline
	2 &          8 &          6 &        1  &    --    &  --   & --\\
	\hline
	3 &        128 &        104 &        22 &        1 &  --   & --\\
	\hline
	4 &      32768 &      26752 &      5736 &      278 &     1 & --\\
	\hline
	5 & 2147483648 & 1753251840 & 375941248 & 18224744 & 65814 & 1 \\
	\hline
	\end{tabular}
}	
\end{table}	
\begin{example}
\label{Ex10}
	Let us consider $f\in {\cal{B}}_4$ whose ANF, the coordinates' (or bits') numbers (these which are greater than 9 are represented by their last digit) and the masks (for $n=4$) are given in Table \ref{Ta3:Example}. When we use the byte-wise WLO Algorithm, it checks consecutively the coordinates of $A_f$, from right to left, i.e., 15, 14, 13, 11, 7, 12---see the WLO sequence $l_4$ in Table \ref{Tab:Results-WLO-Alg}. $A_f$ contains zeros in all coordinates before 12-th, in this coordinate $A_f$ contains one and so the algorithm stops after \textbf{6 checks}. Since 12 is a term of $l_{4,2}$, hence $deg(f)=2$. When the bitwise WLO algorithm is used, it computes the conjunctions: $A_f\wedge m_{4,4}=0$, $A_f\wedge m_{4,3}=0$, $A_f\wedge m_{4,2}>0$ and thereafter it stops. So $deg(f)=2$ and it is computed in \textbf{3 steps}.   
\begin{table}[ht]
  \caption{The data used in Example \ref{Ex10}}
  \label{Ta3:Example}
  \centering
  \small{
  \begin{tabular}{|r|l|}
  \hline
	             Coordinates' numbers & 0 1 2 3 4 5 6 7 \, 8 9 0 1 2 3 4 5 \\
	\hline
	                           $A_f=$ & 1 0 0 1 0 1 1 0 \, 1 0 1 0 1 0 0 0 \\
	\hline
	$\#m_{4,0}=32768,\ m_{4,0}=$      & 1 0 0 0 0 0 0 0 \, 0 0 0 0 0 0 0 0 \\
	\hline
	$\#m_{4,1}=26752,\ m_{4,1}=$      & 0 1 1 0 1 0 0 0 \, 1 0 0 0 0 0 0 0 \\
	\hline
	$\#m_{4,2}=5736,\hfill  m_{4,2}=$ & 0 0 0 1 0 1 1 0 \, 0 1 1 0 1 0 0 0 \\
	\hline
	$\#m_{4,3}=278,\hfill m_{4,3}=$   & 0 0 0 0 0 0 0 1 \, 0 0 0 1 0 1 1 0 \\
	\hline
	$\#m_{4,4}=1,\hfill m_{4,4}=$     & 0 0 0 0 0 0 0 0 \, 0 0 0 0 0 0 0 1 \\
	\hline
	\end{tabular}
}	
\end{table}		
\end{example}

	The second idea for a new bitwise algorithm is to check the bits of $A_f$ in accordance with the WLO sequence. This algorithm will be similar to the byte-wise WLO algorithm and it will have a time complexity of the same type: $O(2^n)$. We discarded this idea because the time complexity of the bitwise WLO algorithm is $O(n.2^{n-6})$ and $n.2^{n-6} < 2^n$ when $6<n<64$. But during the revision of this paper, we noticed that for almost $100\%$ of all $f\in {\cal{B}}_n$, the bitwise WLO algorithm performs $O(2^{n-6})$ checks, whereas the byte-wise WLO algorithm (as well as the new bitwise algorithm) performs $O(n)$ checks. Furthermore, the check of a serial bit of $A_f$ (in accordance with the WLO sequence) needs no more than 5 bitwise operations. Hence the new bitwise algorithm will have a small constant hidden in the $O$-notation. For example, the bitwise WLO algorithm will be better for small $n$ (say $n\leq 8$). But for $n=16$ the bitwise WLO algorithm will perform quite more operations than the new bitwise algorithm. The forthcoming tests will show when and how faster is the new algorithm.
\section{Experimental Results}
\label{Exp_Res}
\begin{figure}[ht]
   \centering
       \scalebox{0.75}{\includegraphics{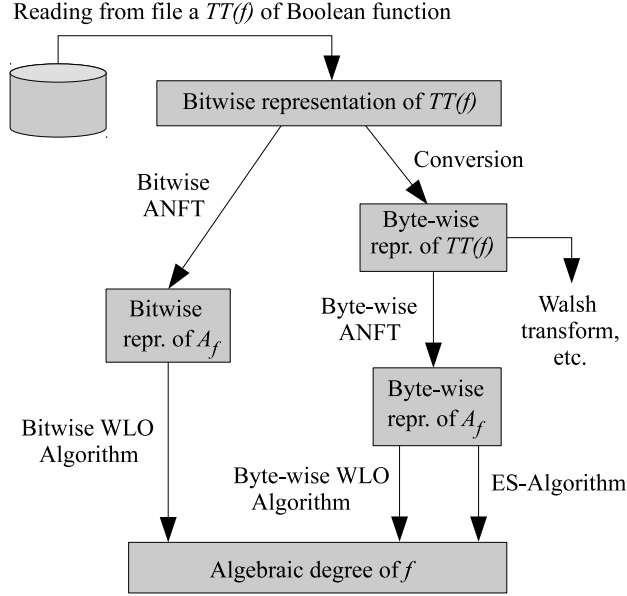}}
   \caption{A scheme for computing the algebraic degree of Boolean functions}
   \label{fig:Scheme}    
\end{figure}

	We return to the main problem of this study---fast computing the algebraic degree of a Boolean function $f \in {\cal{B}}_n$ given by its $TT(f)$. A scheme of the computations and used algorithms is shown in Fig. \ref{fig:Scheme}. 

	In accordance with this scheme, the time complexities of the algorithms considered are summarized as follows: 
\begin{enumerate}
	\item The byte-wise ANFT algorithm followed by the ES algorithm are referred as \textbf{Byte-wise ANFT\&ES} further. So, their time complexity is a sum of $\Theta(n.2^n)+\Theta(2^n)= \Theta(n.2^n)$.
	\item The byte-wise ANFT algorithm followed by the byte-wise WLO algorithm are referred as \textbf{Byte-wise ANFT\&WLO}. Their time complexity is $\Theta(n.2^n)+O(2^n)= \Theta(n.2^n)$. 
	\item The bitwise ANFT algorithm followed by the bitwise WLO algorithm are referred as \textbf{Bitwise algorithms}. When 64-bit computer words are used, the time complexity of the bitwise algorithms is $\Theta((9n-2).2^{n-7}) + O(n.2^{n-6})= \Theta((9n-2).2^{n-7})= \Theta(n.2^n)$. 	
\end{enumerate}

	It has to be noted that these time complexities are:
\begin{itemize}
	\item dominated by the time complexity of the corresponding ANFT---the cost of search is relatively small and it is absorbed into the cost of ANFT;
	\item of the same type $\Theta(n.2^n)$, and the differences between them are in the constants hidden in the $\Theta$-notation.
\end{itemize}

	To understand what these theoretical time complexities mean in practice, we have done a lot of tests. Some more important tests' parameters are:
\begin{enumerate}
	\item Hardware parameters: Intel Pentium CPU G4400, 3.3 GHz, 4GB RAM, Samsung SSD 650 120 GB.
	\item Software parameters: Windows 10 OS and MVS Express 2015 for Windows Desktop. The algorithms are written in C++. All programs were built in Release mode as 32-bit and 64-bit console applications and executed without Internet connection.
	\item Methodology of testing: all tests were executed 3 times, on the same computer, under the same conditions. The running times are taken in average. All results were checked for coincidence. 
The time for reading from file and conversion to byte-wise representation is excluded.
\end{enumerate}

	Table \ref{Ta5:Comparison5vars} shows the obtained running times of the compared algorithms for all $2^{32}$ Boolean functions of 5 variables.
\begin{table}[ht]
  \caption{Experimental results about all $2^{32}$ Boolean functions of 5 variables}
  \label{Ta5:Comparison5vars}
  \centering
  \small{
  \begin{tabular}{|c|c|c|}
  \hline
	Tested     & \multicolumn{2}{c |}{Pure running time in seconds for:} \\
		          \cline{2-3} 
 	algorithms            & 32-bit application& 64-bit application\\
  \hline
    Byte-wise ANFT\&ES  & 540.824      & 507.407 \\
  \hline
    Byte-wise ANFT\&WLO & 450.521	     & 378.374 \\
  \hline
    Bitwise algorithms	& 6.470        & 6.512 \\
  \hline
	\end{tabular}
}	
\end{table}

	Functions of 6 and more variables have been tested with the file of $10^8$ integers. Depending on the number of variables, $2^{n-6}$ integers are read from the file and so they form the serial Boolean function. Tables \ref{Ta4:Comparison_more_vars-32} and \ref{Ta5:Comparison_more_vars-64} show the results for Boolean functions (BFs) of 6 and more variables (vars).

\begin{table}[!ht]
	\caption{Experimental results for 32-bit applications}
  \label{Ta4:Comparison_more_vars-32}
  \centering
  \small{
  \begin{tabular}{|c|c|c|c|c|c|}
  \hline
  \textbf{32-bit} & \multicolumn{5}{c |}{Pure running time in seconds for Boolean functions of:} \\
		  \cline{2-6} 
    implementation  &	6 vars,    & 8 vars,      & 10 vars,      & 12 vars,       & 16 vars, \\
    of:             &	$10^8$ BFs & $10^8/4$ BFs & $10^8/16$ BFs & $10^8/64$ BFs & 97\,656 BFs \\  
  \hline
  Byte-wise ANFT\&ES & 38.834 &  42.400 & 42.664 & 43.466 & 44.740 \\
  \hline
  Byte-wise ANFT\&WLO & 22.003 &  20.022 & 18.758 & 18.230 & 18.808 \\
  \hline
  Bitwise algorithms  &  1.078 &   1.958 &  1.560 &  1.563 &  1.431 \\
  \hline
	\end{tabular}
}
\end{table}

\begin{table}[!ht]
	\caption{Experimental results for 64-bit applications}
  \label{Ta5:Comparison_more_vars-64}
  \centering
  \small{
  \begin{tabular}{|c|c|c|c|c|c|}
  \hline
  \textbf{64-bit} & \multicolumn{5}{c |}{Pure running time in seconds for Boolean functions of:} \\
		  \cline{2-6} 
  implementation      &	6 vars,    & 8 vars,      & 10 vars,      & 12 vars,       & 16 vars, \\
  of:                 & $10^8$ BFs & $10^8/4$ BFs & $10^8/16$ BFs & $10^8/64$ BFs & 97\,656 BFs \\
  \hline
  Byte-wise ANFT\&ES  & 37.429 & 39.178 & 37.699 & 38.789 & 40.350 \\
  \hline
  Byte-wise ANFT\&WLO & 17.443 & 15.880 & 14.224 & 14.243 & 14.454 \\
  \hline
  Bitwise algorithms  &  0.861 &  0.819 &  0.709 &  0.640 &  0.718 \\
  \hline
	\end{tabular}
}
\end{table}

\section{Conclusions}
	We hope that the obtained results show convincingly the advantages of the WLO approaches in computing the algebraic degree of Boolean functions. The bitwise implementations of the considered algorithms are dozens of times faster than the byte-wise implementations. Their usage economizes valuable time, especially in generating S-boxes. The natural continuation of the topic under consideration includes an experimental study of:
\begin{itemize}
	\item The second bitwise algorithm proposed at the end of Section \ref{BitwiseAppr}.
	\item Combination of both approaches discussed in Section \ref{Intro} as follows. First, compute the weight of $TT(f)$. If it is an odd number, then $f$ is of maximal degree. Otherwise, continue with the bitwise algorithms. Some tests with the largest file (of $10^9$ integers) have already begun. The first results show that due to this modification, the bitwise algorithms run about two times faster.
	\item More appropriate software environment (for example, Linux) in order to minimize the effects of background processes running during the executions of the tests. Afterward, repeat all tests since some running times in the last two tables are less than one second and they might not been precise enough.
	\item Application of the bitwise algorithms in computing the algebraic degree of true examples of S-boxes. 
	\item Parallel implementations of the bitwise algorithms.
\end{itemize}
\section*{Acknowledgments}
\label{Acknowlg}
	The author is grateful for the partial support from the Research Fund of the University of Veliko Tarnovo, Bulgaria, under Contract FSD-31-340-14/26.03.2019.

\end{document}